# Vulnerability of the Synchronization Process in the Quantum Key Distribution System

A. P. Pljonkin, Southern Federal University, Taganrog, Russia

## ABSTRACT

A typical structure of an auto-compensation system for quantum key distribution is given. The principle of operation of a fiber-optic system for the distribution of quantum keys with phase coding of photon states is described. The operation of the system in the synchronization mode and the formation of quantum keys was investigated. The process of detecting a time interval with an optical synchronization pulse is analyzed. The structural scheme of the experimental stand of the quantum-cryptographic network is given. Data are obtained that attest to the presence of a multiphoton signal during the transmission of sync pulses from the transceiver station to the coding and backward direction. The results of experimental studies are presented, which prove the existence of a vulnerability in the process of synchronization of the quantum key distribution system. It is shown that the use of a multiphoton optical pulse as a sync signal makes it possible for an attacker to unauthorized access to a quantum communication channel. The experimental results show that tapping a portion of the optical power from the quantum communication channel during the synchronization process allows an attacker to remain unnoticed while the quantum protocol is operating. Experimentally proved the possibility of introducing malfunctions into the operation of the quantum communication system at the stage of key formation, while remaining invisible for control means.

## KEYWORDS

## 1. INTRODUCTION

Modern cryptographic protocols that ensure the security of transmitted messages have a high resistance to burglary. The stability of ciphers is based on mathematical formulations and the limited computing resources of the attacker. It is believed that until now the most reliable security in the transmission of messages provides the use of one-time pads. The development of symmetric methods of encryption is limited to the main problem in the transmission of confidential information, which is formulated as the problem of distributing a secret key between legitimate users.

The well-known Shannon rule, which interprets the use of a secret key for a secure transmission, is updated with the development of new technologies for the formation of secret keys. Thus, the achievement of absolute secrecy in the transmission of messages is possible only by solving the problem of key distribution.

The development of methods of quantum cryptography to ensure security in telecommunications systems of information transmission theoretically allows to achieve absolute secrecy of ciphers (Gisin et al., 2002). Quantum cryptography is based on the laws of quantum physics and is based on the coding of the quantum state of a single particle. The essence of quantum cryptography lies in the reliable distribution of the secret key between legitimate users. Another component in the quantum distribution is the creation of a random secret key (Bennet et al., 1992; Stucki et al., 2002; Broadbent & Schaffner, 2007).

Practical implementation of quantum cryptography is based on quantum key distribution systems (QKDS). If the existing encryption algorithms can be distorted by mathematical improvements, then quantum cryptography is the only way to solve the problem of key distribution. Recall that the basis of quantum cryptography lies in the following statements: it is impossible to clone an unknown quantum state and it is impossible to obtain information on non-orthogonal quantum states without perturbation. Consequently, any unauthorized measurement will lead to a change in the quantum state.

In quantum cryptography, symmetric cryptosystems are common (Makarov, 2007). In such systems, one key is used for both encryption and decryption. Messages sent along the lines of quantum communication, theoretically can't be intercepted or copied. Quantum key distribution is a technology based on the laws of quantum physics to create a sequence of random bits in two remote users. This sequence is used as a cryptographic key, and the key array itself is called a "one-time pad.

## 2. QUANTUM KEY DISTRIBUTION SYSTEMS

In 2007, the methods of quantum cryptography were first applied in a large-scale project. Quantum security system, developed by the Swiss company idQuantique, was used to transmit voting data at the parliamentary elections in Geneva. To date, really functioning quantum communication systems have been created. The efforts of developers are now aimed at increasing the communication range, increasing the speed of forming a quantum key, improving the characteristics of fiber-optic components.

As noted earlier, a symmetric cryptosystem generates a shared secret key and distributes it among legitimate users to encrypt and decrypt messages (Rumyantsev & Pijokin, 2015) . An attacker attempting to investigate transmitted data can't measure photons without distorting the original message. The system on the open channel compares and discusses signals transmitted on the quantum channel, thereby verifying them for the possibility of interception. If the system does not contain errors, then the transmitted information can be considered securely distributed and secret, despite all the technical capabilities that a cryptanalyst can use.

Quantum key distribution systems operate under the control of quantum protocols. There are several protocols of quantum cryptography based on the coding of single photon states, for example: BB84, B92, Koashi-Imoto, SARG04 and their modifications (Kurochkin et al., 2012). Under the signal in quantum communication systems is meant the transmitted quantum state of a photon. The first protocol that was implemented in the QKD systems is called BB84. The basis of the BB84 protocol is the principles of particle phase coding and auto-compensation of polarization distortions. This protocol is also called bi-directional because of the propagation of the optical signal along a single fiber-optic path in two directions. Note that today the BB84 protocol has more efficient modifications. In the known BB84 protocol, the receiver analyzes the photons and randomly selects the polarization measurement method. On an unprotected channel, the receiver informs the sender of the method of choosing the basis for each photon, without revealing the measurement results themselves. After that, the sender on an unprotected channel tells you whether the type of measurement for each photon is correctly selected. As a result, an unrefined (raw) key is generated.

## 3. SYNCHRONIZATION IN QKD SYSTEM

The QKD system can't operate without synchronization (Lydersen et al., 2010). During the synchronization process, optical pulses propagate from the transceiver station to the encoding and vice versa. The synchronization task is the detection of the signal time interval with maximum accuracy.

We will explain that under the signal time interval is meant a time window containing an optical sync pulse. The pulses are recorded by one-photon avalanche photodetectors of the transceiver station. Let us consider in more detail the synchronization process using the example of an active quantum key distribution system with phase coding of photon states. The process consists of three stages, each of which is a continuation of the previous one and consists in determining the moment of detection of the optical pulse by single-photon photodetectors. The problem of detecting the signal time interval is solved by measuring the propagation path length of the optical pulse from the transceiver station to the encoding and vice versa. The structural scheme of the synchronization process of the auto-compensation QKD system is shown in Figure 1.

A laser diode generates optical pulses at a wavelength of 1550 nm and a duration of about 1 ns. The period and duration of the optical pulse are absolutely stable. The repetition period $T_s$ is determined by the length of the quantum communication channel between the stations of the QKD system. The time frame equal to the optical pulse repetition period $T_s$ is divided into $N_w$ of time windows with a duration $t_w$ so that $T_s = N_w \bullet t_w$. Photodetectors are put into working mode and a sequential polling of all temporary windows begins. Each window is analyzed N times. Thus, the value of N is formulated as the sample size in the time window. When analyzing each time window, the number of registered photoelectrons (PEs) and / or pulses of dark current (PDC) is fixed. After polling all $N_w$ time windows, an array of values of registered PE and / or PDC is formed. The conditions for detecting the optical pulse by the detection equipment during synchronization are described in [9]. The time window with the maximum number of registered FEs is recognized as a signal window. At each next stage of synchronization, the duration of the time window is reduced. So, in the first stage $t_w = 300ps$, on the second $t_w = 60ps$, at the third stage $t_w = 10ps$.

Note that the accuracy of the time window in 10ps makes it impossible to change the physical length of the quantum communication channel after the synchronization process. The latter requires periodic initialization of the synchronization process under real conditions of operation of quantum distribution systems, since even a slight atmospheric effect on the quantum channel causes deformation of the optical fiber.

To study the process of entering synchronism, the experimental stand is assembled (Figure 2). Two stations of the QKD system are interconnected by a fiber-optic communication line. Each station is controlled by software. To construct an energy model of the quantum key distribution system, six optical control points measure the power level of the optical radiation. The connections were made using plug-in connections. The measurements were carried out using the Yokogawa optical modular system and the LeCroy digital oscilloscope.

The energy model of the system made it possible to calculate the power loss of optical pulse on all sections of the fiber-optic signal propagation path (Pijokin & Rumyantsev, 2016). The latter shows that during the synchronization (detection of a time interval with an optical pulse), the signal optical pulse contains more than $10^3$ photons in back propagation from the coding station to the transceiver. Note that the transmission of synchronizing signals from the transceiver station to the coding one is always performed in multiphoton mode. In addition, it is established that avalanche photodiodes function in the linear mode during synchronization, and the procedures for error correction and power control (as in the operation of the quantum protocol) are completely absent.

Thus, it can be concluded that the synchronization process in quantum key distribution systems with automatic compensation of polarization distortions takes place in a multiphoton mode and does not have means of protection against unauthorized access.

**Figure 1. Sync pulse detection. Ld – laser diode; SPADs – single-photon avalanche photodiodes; Dl – delay line of optical radiation; BPS – beam polarizing splitter; Od – optical divisor; Fm is a Faraday mirror; QC – quantum channel; Delay - time delay of gating; Nw1 (Nwn) – the number of time windows; Counts – the moment when photodetectors are triggered; tw – duration of the time window.**

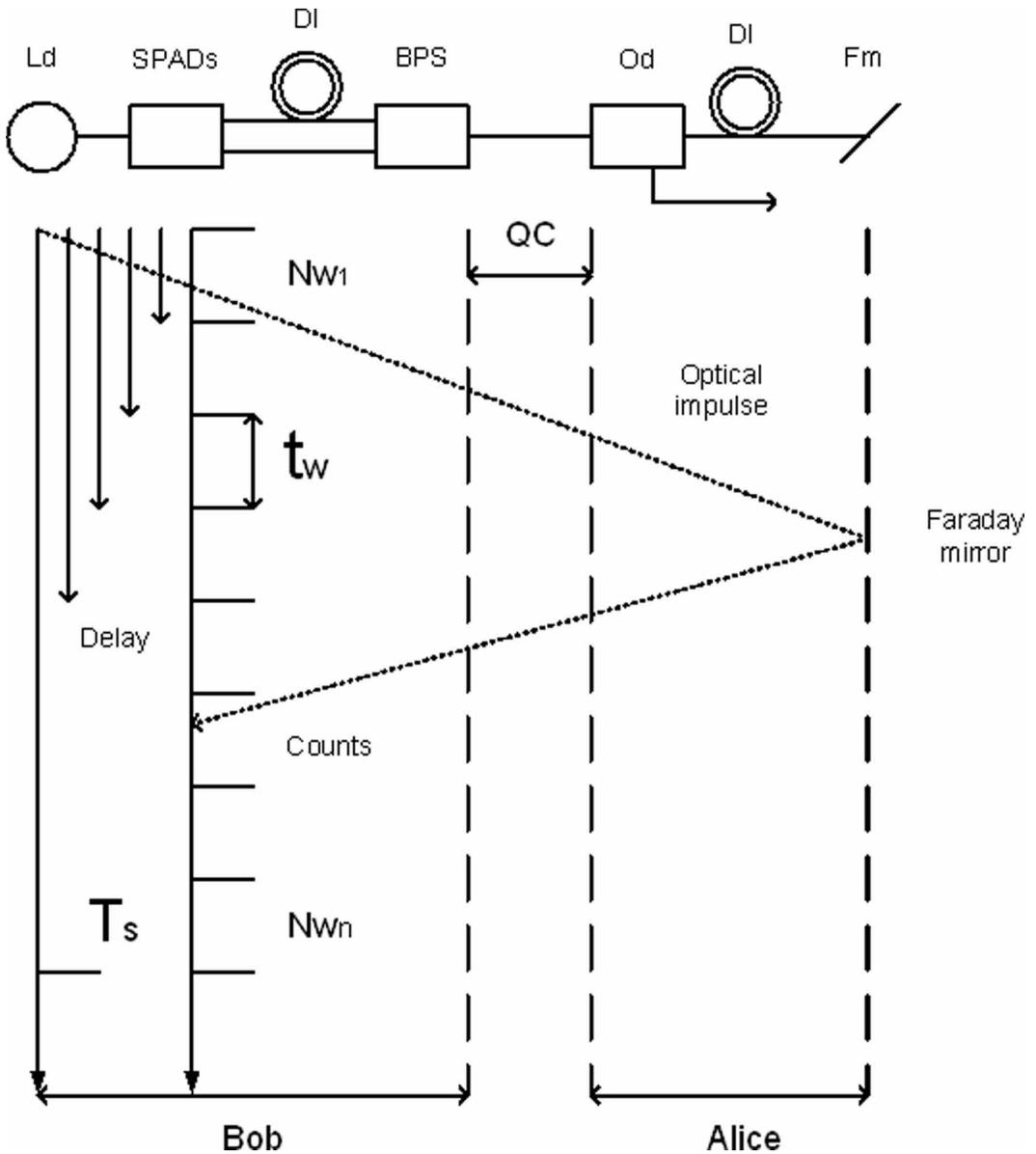

## 4. ATTACK ON THE QUANTUM CHANNEL EXPERIMENTS

Note that back in 2007, a group of scientists successfully carried out an attack on the quantum key distribution system with phase coding (Rumyantsev & Pijokin, 2016). The attack was based on imperfection of the system and was aimed at destabilizing the quantum protocol. During the implementation of the attack, it was assumed that the system was already synchronized.

Let us prove that the vulnerability of the synchronization process can be used to further interfere with the operation of the quantum key distribution system at the stage of functioning of the quantum protocol.

**Figure 2. Structural scheme of the optical part of the QKD system. Ld – laser diode; C – optical circulator; Od – directional coupler; SPAD – photodetectors; Module B – small radiation delay line, phase modulator, filter; BPS – beam polarizing splitter; QC – quantum channel; At – attenuator; G – circuit with clock generators; DI - radiation delay line with a length of 24 km; Pm - phase modulator; Fm – Faraday mirror; S – medium converter; P.Meter – optical power meter.**

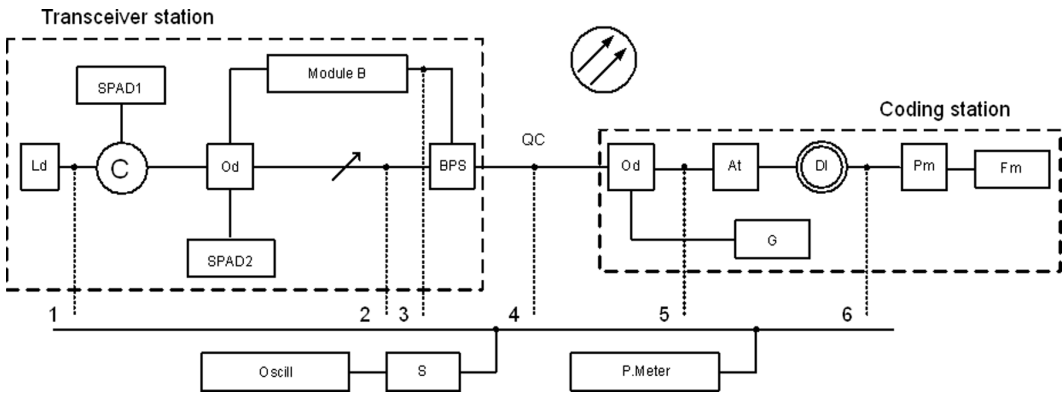

Multiphoton mode potentially allows an attacker to organize unauthorized access to a fiber-optic communication channel (to a quantum channel). The purpose of unauthorized access can be not only the interception and reading of information, but also the synchronization of the equipment of the attacker with the aim of interfering with the operation of the QKD system. Attack "Trojan horse" is an example of the closest to practical implementation, when it is possible to replace the original messages with an attacker's messages. The attacker connects to the quantum communication channel and generates copies of optical pulses intercepted from the transceiver station and then sent to the encoding station. Thus, legitimate users can not recognize the presence of an attacker in a quantum communication channel. Realized auto-compensation systems with phase coding of photon states function according to a two-pass scheme, i.e. optical signals propagate along a single fiber in both directions. Such a realization complicates the problem of unauthorized access to a quantum communication channel but does not completely exclude it. The moment of interception of the optical pulse during direct propagation of the signal does not give complete information to the attacker about the operation of the system. The decisive moment is the appearance of an optical pulse in the backward propagation of a reflected signal in a quantum channel. Having information about the time of re-reflection, the attacker is able to simulate the work of the encoding station and at the right time send imitation signals to the photodetectors of the transceiver station.

Figure 3 (a) shows the scheme of the experiment with an optical power coupler integrated into the quantum communication channel. Coefficients of division 90/10 (%). 10% of the power of optical radiation is diverted to the measuring equipment with a direct signal passing (from the transceiver station to the coding one). 90% are sent to the coding station. At backward propagation the signal without losses arrives at the transceiver station. With this modification, not only the synchronization process, but also the process of quantum key distribution (quantum protocol) function in the regular mode. The theoretical error, which is calculated by the software of the QKD system, was 2.85%. The actual error was 0.65%. Note that the values obtained are not critical; the system does not detect the removal of part of the optical power from the quantum communication channel.

Let us realize the problem of removing part of the optical power at two points of the quantum communication channel. For this purpose, a circuit was constructed using two directional fiber-optic couplers with division coefficients of 85/15 (%) and 70/30 (%). The couplers are integrated in the quantum communication channel as shown in Figure 3 (b).

To prove the thesis that the optical fiber elements of the coding station are inactive during the synchronization process, we will conduct the following experiment: during the detection of the length of the fiber-optic communication line, we detach the transceiver station from the quantum channel and give the optical pulse 0dBm to the input of the 85/15 coupler in the direction of the coding station. At the same time, we capture the values at the outputs of the couplers 15% and 30% – -9.6 dBm and -57.7 dBm, respectively. We repeat the measurements, but the coding station is de-energized this time. The values at the outputs of the couplers of 15% and 30% are -9.6 dBm and -57.7 dBm, respectively. Thus, the experimental results show that the active fiber-optic elements of the coding station are not involved in the synchronization process and do not affect the optical signal.

**Figure 3. (a) use of a directional coupler in a quantum communication channel; (b) extraction of optical power at two points**

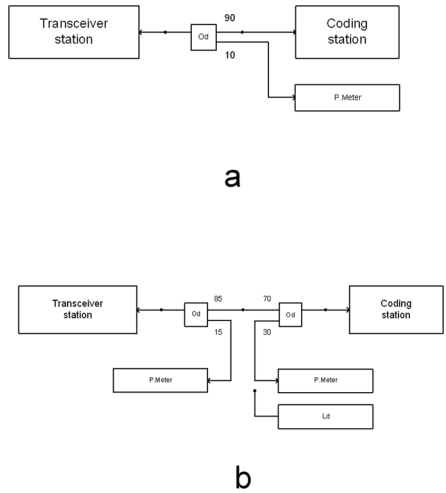

As noted earlier, the attacker's task may be not only to intercept and decrypt information transmitted via the quantum communication channel, but also to introduce interference to destabilize the system. We will experimentally prove the possibility of introducing interference into the operation of the quantum protocol, without detecting the presence of an attacker in the quantum communication channel.

In accordance with the scheme presented in Figure 3 (b), we will integrate fiber-optic couplers into a fiber-optic communication line between two stations of the QKD system. The first stage of the experiment is organized using a quantum communication channel with a length of 1000 m. The operation of the quantum key distribution system is started in the regular mode. According to the algorithm of work, the system subsequently tests the laser diode and photodetectors, analyzes the value of losses in the fiber-optic communication line. After the system analysis, the process of detecting the length of the quantum communication channel (synchronization) is started. The system synchronized without detecting the presence of two couplers in the quantum communication channel. Then the key distribution process is initiated. In fact, the measured error was 2.97%. The operation of the quantum protocol also did not detect the presence of couplers. The circuit with two integrated couplers operated continuously for 36 hours. The key distribution and synchronization process worked in a cyclic mode, the keys were accumulated in the buffer. At the second stage of the experiment, during the distribution of the quantum keys, a source of radiation with a wavelength of 1550 nm with a power of about 1 mW and a repetition rate of 270 Hz was connected to the output of the divider 15%. The level of the measured error increased from 2.97 to 3.35%, but the process of forming the quantum keys was not violated and continued normal operation. Next, we connected a radiation source with the same parameters to the output of 30%, while the pulses were sent to the coding station. With such a scheme, the source of radiation in the key formation mode was briefly switched on. Note that in the presence of two couplers with respect to signal parameters, you can determine the mode in which QKD systems work without difficulty. Because of the impact on the coding station on the assembled circuit, the process of forming the quantum key ceased to function. The system entered the tuning mode without detecting the presence of couplers (the error level in the system did not change). Violation of the process of key formation is perceived by the system not as an attacker in the communication channel, but as a mismatch of the frame structure of the sync pulse frames. After turning off the source of interference, the system is adjusted and continues to operate the quantum protocol. The experiment was repeated for a length of a quantum communication channel of 2, 4, and 6 km according to a similar scheme. The results of the experiment remained unchanged, for all fiber-optic communication lines the system did not detect the presence of two couplers in the optical communication channel.

We note that similar studies in related fields are also relevant (Rumyantsev & Pijokin, 2015; Pijonkin, 2017; Pijokin et al., 2017; Yuen, 2016; Pijonkin et al., 2016; Distribution, 2010; Chan et al., 2011; Advanced in Security and Privacy…, 2018; Botnet-based distributed denial of service…, 2012; AIZain et al., 2015; Gupta et al., 2016; Gupta et al., 2017; Bushan et al., 2017; Gupta et al., 2017; Aakanksha et al., 2017; Shashank et al., 2017).

## 5. CONCLUSION

Thus, it is proved that if an intruder enters the quantum communication channel at the stage of configuring a QKD system, the attacker may remain unnoticed during synchronization and during the formation of the quantum key. The latter makes it possible to interfere with the operation of the QKD system without revealing its presence. The results of the experiment show that it is not necessary for an attacker to have expensive equipment for intercepting and decrypting quantum keys. The presence of standard optical power couplers and access to the fiber-optic communication lines between the stations of the QKD system allows to interfere with the system operation at the required time, while remaining unnoticed.

The results of experimental studies allow us to formulate concrete conclusions about the vulnerability of the synchronization process of the QKD system: during the synchronization, the active nodes of the coding station are inactive and, therefore, the algorithms for ensuring the protection and control of the optical sync pulse are not functioning; the transmission of optical signals is carried out in multiphoton mode, which allows an attacker to remove a part of the optical power from the quantum communication channel, remaining undetected; an attacker has enough typical fiber optic equipment to interfere with the operation of the system at the stage of quantum key distribution, while remaining unnoticed for control.

## ACKNOWLEDGEMENT

Work is performed within the grant of President of Russian Federation for state support of young Russian scientists MK-2338.2018.9 "Creation of an automated algorithm for integrating quantum keys into the data network while providing enhanced security against unauthorized access to the quantum communication channel".

*Pljonkin (Plenkin in EU) Anton Pavlovich b. 1984 Web of Science: L-3806-2016 Scopus: 57190492977 Orcid: 0000-0001-6713-9347 ResearcherID: L-3806-2016 Associate Professor Master of Science, Ph.D. Research laboratory "Quantum cryptography" Southern Federal University pljonkin@mail.ru +79054592158 Skype: pljonkin FB: Pljonkin Anton*